\documentclass[%
 reprint,
superscriptaddress,
 amsmath,amssymb,
 aps,
]{revtex4-1}

\usepackage{graphicx,setspace}
\usepackage{dcolumn}
\usepackage{bm}
\usepackage[mathlines]{lineno}
\usepackage[colorlinks=true,citecolor=blue,linkcolor=blue,anchorcolor=green, anchorcolor=green,urlcolor=blue]{hyperref}
\usepackage{mathrsfs}
\usepackage{xcolor}
\usepackage{float}
\usepackage{ulem}
\usepackage{lipsum} 

\let\originalleft\left
\let\originalright\right
\renewcommand{\left}{\mathopen{}\mathclose\bgroup\originalleft}
\renewcommand{\right}{\aftergroup\egroup\originalright}
\mathcode`\*="8000
{\catcode`\*=\active\gdef*{\mathclose{}\,\mathopen{}}}

\newcommand{\cu}[1]{\left\{#1\right\}}

\newcommand{\be}{\begin{equation}}
\newcommand{\ee}{\end{equation}}
\newcommand{\bea}{\setlength\arraycolsep{2pt} \begin{eqnarray}}
\newcommand{\eea}{\end{eqnarray}}
\newcommand{\nn}{\nonumber}

\begin{document}

\title{Observational signature of a near-extremal Kerr-Sen black hole in the heterotic string theory}

\author{Minyong Guo}
\email{minyongguo@pku.edu.cn}
\affiliation{Center for High Energy Physics, Peking University, No.5 Yiheyuan Rd, Beijing 100871, China.}

\author{Shupeng Song}
\email{songsp@mail.bnu.edu.cn}
\affiliation{Department of Physics, Beijing Normal University,
Beijing 100875, China.}
\affiliation{Institute for Gravitation and the Cosmos \& Physics Department, Penn State, University Park, PA 16802, USA.}

\author{Haopeng Yan}
\email{Corresponding author:haopeng.yan@nbi.ku.dk}
\affiliation{\small
The Niels Bohr Institute, University of Copenhagen, Blegdamsvej 17, 2100 Copenhagen {\O}, Denmark.}

\begin{abstract}
  We analytically study the optical appearance of an isotropically emitter orbiting near the horizon of a near-extremely rotating Kerr-Sen (KS) black hole which is an electrically charged black hole arising in heterotic string theory. We study the influence of the Sen charge on the observational quantities, including the image position, flux and redshift factor. Moreover, we compare the results with those for a near-extremal Kerr-Newman (KN) black hole, which is the charged rotating black hole in general relativity. We find quantitative corrections of the signatures of these charged black holes (both KS and KN) compare to that of a neutral Kerr black hole. This may serve as distinctive features of different black holes for future tests by the Event Horizon Telescope.
\end{abstract}

\maketitle

\section{Introduction}
\label{sec:intro}
Black holes are among the most important predictions of general relativity (GR), as well as other gravitational theories. Thus, attempts to discover black holes have received continuing impetus over many decades. Now it eventually has become reality with the detections of gravitational waves (GWs) by LIGO and Virgo \cite{Abbott:2016blz,Abbott:2016nmj} and with the first image of the black hole M87 photographed by the Event Horizon Telescope (EHT) Collaboration \cite{Akiyama:2019cqa,Akiyama:2019brx,Akiyama:2019sww,Akiyama:2019bqs,
Akiyama:2019fyp,Akiyama:2019eap}. Yet, there is still an urgent need for more precise theoretical templates to match these data,
which has triggered exciting research among the gravity community \cite{Barack:2018yly,Cunha:2018acu,Gralla:2019xty,Johnson:2019ljv,Bambi:2019testing}.

From the first image of the M87, a bright ring surrounding a dark region was observed as an important feature of a black hole. The dark region is known as the ``black hole shadow" and the bright ring is elicited by the luminous sources outside the black hole \cite{Akiyama:2019cqa}. This observed bright ring has an intricate substructure, which can produce strong and universal signatures on long interferometric baselines \cite{Johnson:2019ljv}. As we know, a black hole is invisible and it is the surrounding luminous matters that make the black hole observable. Thus, the appearance of a black hole depends closely on its surroundings. Within the surrounding luminous matter, a bright point emitter (localized emissivity enhancement, also refer to as ``hot spot") is particularly interesting and can produce striking observational signals. In the 1970s, the optical appearance of a hot spot (star) orbiting on a circular orbit of the extremal Kerr black hole has been studied in Refs.~\cite{cunningham1972optical,cunningham1973optical}. Recently, the observational signature produced by a hot spot on, or near, the innermost stable circular orbit (ISCO) of a near-extremal Kerr black hole has been studied in Ref.~\cite{Gralla:2017ufe}, where a striking signature of high spin Kerr black hole was found. Later, the influence of a surrounding plasma on that signature has been studied in Ref.~\cite{Yan:2019etp}. Besides these, the image of a non-stationary plunging hot spot approaching a black hole has been studied in Refs.~\cite{Dokuchaev:2018kzk,Dokuchaev:2019bbf,Dokuchaev:2019pcx}. All these studies are based on the assumption that the underlying gravity theory is general relativity. Nevertheless, there are also black holes in alternative gravity theories \cite{sen1992rotating,ayzenberg2014slowly,
myers1986black,amarilla2010null,abdujabbarov2013shadow,
moffat2006scalar} based on different motivations. It is interesting to study the observational signature of a black hole (and hot spot) in these theories. Given this motivation, the signature of a Kerr-MOG (KM) black hole in the scalar-tensor-vector modified gravity theory (MOG) has been studied in Ref.~\cite{Guo:2018kis}. Among other gravity theories that modify GR, perhaps the most theoretically important one is string theory, since it is one of the most attractive candidates for quantum gravity and unified theory. Thus, in this paper we will study the observational signals produced by a hot spot orbiting a near-extremal rotating black hole arising from the string theory.

In a low-energy limit of heterotic string theory, a rotating black hole solution has been found in Ref.~\cite{sen1992rotating}, known as the Kerr-Sen (KS) black hole. In addition to the mass $M$ and angular momentum $J$, the KS black hole has a third physical parameter: the electric charge $Q$ corresponding to a $U(1)$ gauge field. Much attention has been paid to the KS black hole including studies of null geodesics, photon motion and optical appearance (shadow) of a black hole \cite{An:2017hby,Gyulchev:2006zg,Younsi:2016azx,PhysRevD.78.044007,Dastan:2016bfy,Uniyal:2017yll}. 
Moreover, the apparent shape in the KS spacetime has also been compared to those in various charged/rotating black holes and naked singularities of the Kerr-Newman (KN) class of spacetimes \cite{PhysRevD.78.044007}. It is instructive to compare the results for the KS black hole with those for the KN black hole since the later is the charged and rotating solution in the Einstein-Maxwell theory [i.e., GR coupled to a $U(1)$ gauge field]. The main target of this paper is therefore to compute the signature produced by a hot spot near a near-extremal KS black hole, and compare this to that of a near-extremal KN black hole.
By comparison, we find that the expressions of the metrics for extremal KN and KM black hole are mathematically identical upon replacements of two corresponding parameters, and so do the observational quantities. We then quote relevant results for the KM case from Ref.~\cite{Guo:2018kis} and transfer them to the observational quantities for the KN case.
We find that the general qualitative features of the hot spot image in the KS case are the same as those in the Kerr spacetime, while quantitative corrections appear when the Sen charge is non-zero. Moreover, the $U(1)$ charges in both KS and KN cases trend to have positively correlated influences on most of the observables, however, the magnitudes are distinguishable.

This paper is organized as follows. In Sec.~\ref{sec:KerrSenReview}, we review the Kerr-Sen spacetime and the geodesics in this spacetime. In Sec.~\ref{sec:OrbitingEmitter}, we set up the problem of the observational appearance of an orbiting emitter for the general case. We write down the lens equations to be solved and write down the observational quantities that we are interested in. In Sec.~\ref{sec:NearExtremalExpansion}, we solve this problem for the near-extremal case to the leading order in the deviation from extremality. In Sec.~\ref{sec:ResultsDiscussion}, we present our results with a figure and compare the results with those for the KN case with a table, and we discuss these in detail. In Appendix~\ref{app:KN}, we revisit the KN spacetime and introduce its observational appearance. In the remaining Appendixes, we describe some technical steps.
Unless otherwise stated, we set $c=G_N=1$ throughout the paper.


\section{Kerr-Sen black hole and geodesics in Kerr-Sen spacetime}
\label{sec:KerrSenReview}
The Kerr-Sen (KS) black hole spacetime is described by a 4-dimensional effective action arising from heterotic string theory:
\begin{eqnarray}
\label{action}
S=-\int d^{4}x\,\sqrt{-\mathcal{G}}e^{-\Phi}\bigg(&&-\mathcal{R}+\frac{1}{12}
\mathcal{H}^{2}\nn\\
&&-\mathcal{G}^{\mu\nu}\partial_{\mu}\Phi\partial_{\nu}\Phi
+\frac{1}{8}\mathcal{F}^{2}\bigg),
\end{eqnarray}
where $\mathcal{R}$ and $\Phi$ are the scalar curvature and the dilation field, respectively, and $\mathcal{F}^{2}=\mathcal{F}_{\mu\nu}\mathcal{F}^{\mu\nu}$ with $\mathcal{F}_{\mu\nu}$ being the field strength corresponding to the Maxwell field $\mathcal{A}_{\mu}$, and
\begin{eqnarray}
\mathcal{H}^{2}=\mathcal{H}_{\mu\nu\rho}\mathcal{H}^{\mu\nu\rho}.
\end{eqnarray}
Here, the expression of $\mathcal{H}_{\mu\nu\rho}$ is of the form
\begin{eqnarray}
  \mathcal{H}_{\mu\nu\rho}&=&\partial_{\mu}\mathcal{B}_{\nu\rho}
                 +\partial_{\nu}\mathcal{B}_{\rho\mu}
                 +\partial_{\rho}\mathcal{B}_{\mu\nu}\nn\\
                 &&-\frac{1}{4}\bigg(\mathcal{A}_{\mu}\mathcal{F}_{\nu\rho}
                 +\mathcal{A}_{\nu}\mathcal{F}_{\rho\mu}
                 +\mathcal{A}_{\rho}\mathcal{F}_{\mu\nu}\bigg),\label{Hmunurho}
\end{eqnarray}
with $\mathcal{B}_{\mu\nu}$ being an axion field.
$\mathcal{G}_{\mu\nu}$ appearing in Eq.~(\ref{action}) are the covariant components of the metric in the
string frame, which are related to the Einstein metric by
$g_{\mu\nu}=e^{-\Phi}\mathcal{G}_{\mu\nu}$. The Einstein metric, gauge field strength and electromagnetic potential in Boyer-Lindquist coordinates read \cite{sen1992rotating}:
\bea
\label{metric}
ds^2&=&-\frac{\Delta}{\rho_b^2}(dt-a\sin^2\theta d\phi)^2+\frac{\sin^2\theta}{\rho_b^2}(adt-\delta d\phi)^2\nn\\
&&+\frac{\rho_b^2}{\Delta}dr^2+\rho_b^2d\theta^2\\
&=&-\frac{\Delta\rho_b^2}{\Xi}dt^2+\frac{\rho_b^2}{\Delta}dr^2
+\rho_b^2d\theta^2+\frac{\Xi\sin^\theta}{\rho_b^2}(d\phi-\omega dt)^2,\nn\\
\textbf{H}&=&-\frac{2ba}{\rho_b^2}dt\wedge d\phi\wedge\big[(r^2-a^2\cos^2\theta)\sin^2\theta dr \nn\\
&&-r\Delta\sin2\theta d\theta\big],\nn\\
\textbf{A}&=&-\frac{Qr}{\rho_b^2}(dt-a\sin^2\theta d\phi),\,\,\,\,\,\,\Phi=2\log\left(\frac{\rho}{\rho_b}\right),\nn
\eea
where we have defined
\bea
\label{eq:MetricPara}
\begin{aligned}
\rho^2=r^2+a^2\cos^2\theta,\quad\,\,&\rho_b^2=\rho^2+2br,\\
\delta=r^2+a^2+2br,\quad\,\,\,\,\,&\Delta=\delta-2Mr,\\
\Xi=\delta^2-\Delta a^2\sin^2\theta,\quad &\omega=2aMr/\Xi.
\end{aligned}
\eea
This Einstein metric describes a KS black hole with mass $M$, $U(1)$ charge $Q=\sqrt{2Mb}$, and angular momentum $J=Ma$ with $a$ and $b$ the spin and twist parameters, respectively. The metric reduces to the Kerr geometry when the twist parameter $b$ goes to zero. The horizons are determined by the roots of $\Delta=0$, as
\bea\label{ho}
r_{\pm}=M-b\pm\sqrt{(M-b)^2-a^2},
\eea
where $r_\pm$ denote the outer and inner horizon, respectively. The regularity of the event horizons requires
\bea
\label{eq:regulrity}
a\leq M-b,
\eea
and the extremal case is obtained for equality in \eqref{eq:regulrity} being taken.
When $b$ and $a$ are bounded, respectively, we obtain the corresponding ranges for $a$ and $b$ as
\bea
0\leq a\leq M,\,\,\,\,\,\,\,\,\,\,\,0\leq b\leq M.
\eea

Consider a particle (or photon) of mass $m$ moving in the KS spacetime with 4-momentum $p^a=\left(\frac{\partial}{\partial\tilde{\tau}}\right)^a$ with
$\tilde{\tau}$ an affine parameter which is related to the particle's proper time $\tau$ by $\tilde{\tau}=\tau/m$. Then
the 4-momentum of the particle takes the general form
\bea
p^a=\dot t\left(\frac{\partial}{\partial t}\right)^a+\dot r\left(\frac{\partial}{\partial r}\right)^a+\dot \theta\left(\frac{\partial}{\partial \theta}\right)^a+\dot \phi\left(\frac{\partial}{\partial \phi}\right)^a,
\eea
where ``\,$\cdot$\," denotes the derivative with respect to $\tilde{\tau}$.
The conserved quantities along the trajectory are
\begin{subequations}
\bea
-m^2&=&g_{ab}p^ap^b,\\
E&=&-g_{ab}p^a\left(\frac{\partial}{\partial t}\right)^b=-p_t,\\
L&=&g_{ab}p^a\left(\frac{\partial}{\partial \phi}\right)^b=p_\phi,\\
\mathcal{Q}&=&p_\theta^2-\cos^2\theta(a^2p_t^2-p_\phi^2\csc^2\theta)
+m^2a^2\cos^2\theta,\,\,\,\,\,\,\,\,\,\,
\eea
\end{subequations}
where $E$ is the total energy, $L$ is the angular momentum and $\mathcal{Q}$ is the Carter constant \cite{carter1968global}.
Using the Hamilton-Jacobi method, we get the 4-momentum up to two choices of sign corresponding to the direction of the trajectory,
\begin{subequations}
\label{ge}
\bea
\Delta p^t&=&\frac{\delta}{\Delta}(E\delta-aL)-a(aE\sin^2\theta-L),\\
\Delta p^r&=&\pm\sqrt{\tilde{\mathcal{R}}(r)},\\
\Delta p^\theta&=&\pm\sqrt{\tilde{\Theta}(\theta)},\\
\Delta p^\phi&=&\frac{a}{\Delta}(E\delta-aL)-\csc^2\theta(aE\sin^2\theta-L),
\eea
\end{subequations}
where
\begin{subequations}
\label{eq:generalPoten}
\bea
\label{eq:radialPoten}
\tilde{\mathcal{R}}(r)&=&(E\delta-aL)^2-\Delta\left[m(r+2br)
+\mathcal{Q}+(L-aE)^2\right],\,\,\,\,\,\,\,\,\,\\
\label{eq:angularPoten}
\tilde{\Theta}(\theta)&=&\mathcal{Q}-ma^2\cos^2\theta-(L^2\csc^2\theta-a^2E^2)
\cos^2\theta.
\eea
\end{subequations}
The functions $\tilde{\mathcal{R}}(r)$ and $\tilde{\Theta}(\theta)$ are
are respectively the radial and angular potentials, the vanishing of which correspond to the radial and angular turning points in the trajectories respectively.

\section{Orbiting emitter near Kerr-Sen black hole}
\label{sec:OrbitingEmitter}
We are interested in an isotropic emitter orbiting on a circular and equatorial geodesic at radius $r_s$ around a KS black hole. For such an emitter, we have $\theta=\pi/2$, $\tilde{\mathcal{R}}(r)=0$ and $\mathrm{d}\tilde{\mathcal{R}}(r)/\mathrm{d}r=0$. Solving these equations simultaneously for $E$ and $L$ gives
\begin{subequations}
\label{eq:ConservedQuantities}
\bea
\label{energy}
\frac{E_{\pm}}{m}&=&\frac{(r+b)^{1/2}(r-2M+2b)\pm a M^{1/2}}{(r+2b)^{1/2}P^{1/2}},\\
\label{angmom}
\frac{L_{\pm}}{m}&=&\frac{\pm M^{1/2}[r(r+2b)\mp 2a M^{1/2}(r+b)^{1/2}+a^2]}{(r+2b)^{1/2}P^{1/2}},\,\,\,\,\,\,\,\,\,\,\label{angmom}
\eea
\end{subequations}
where
\be
P=r^2-(3r+2b)(M-b)\pm 2a M^{1/2}(r+b)^{1/2}.
\ee
Note that the existence of circular orbits requires that the denominator of \eqref{energy} and \eqref{angmom} is real, i.e.,
\be
(r+2b)P>0.
\ee
Combining the equations \eqref{ge} and \eqref{eq:ConservedQuantities} we obtain the angular velocity of the emitter, as
\begin{equation}
\label{eq:AngularVelocity}
\Omega_s\equiv\frac{d\phi}{dt}
=\frac{\pm M^{1/2}}{(r+2b)(r+b)^{1/2}\pm a M^{1/2}},
\end{equation}
where the plus and minus sign denote the direct or retrograde orbits, respectively. Here and hereafter, we use the subscript $s$ to represent the \lq\lq{}source\rq\rq{}.

\subsection{Photon motion and lens equations}
\label{subsec:PhotonMotion}
For a photon trajectory, we have $m=0$ in the geodesic equations~\eqref{ge}. In this case, the energy $E$ may be scaled out from these equations and it is convenient to introduce two new dimensionless parameters
\bea
\label{hatlambdaQ}
\hat\lambda=\frac{L}{E},\,\,\,\,\,\,\,\,\hat q=\frac{\sqrt{\mathcal{Q}}}{E},
\eea
to describe this photon trajectory.
Note that we will always have real $\hat{q}>0$ for all trajectories that intersect the equatorial plane since $\mathcal{Q}=p_\theta^2\geq0$.
In terms of the new parameters, the potentials \eqref{eq:generalPoten} for the null case can be written as
\begin{subequations}
\bea
\label{eq:NullRadialPoten}
\mathcal{R}(r)&=&\frac{\tilde{\mathcal{R}}(r)}{E^2}=(a\hat\lambda-\delta)^2-\Delta\left[\hat q^2+(\hat\lambda-a)^2\right],\\
\label{eq:NullAngularPoten}
\Theta(\theta)&=&\frac{\tilde{\Theta}(\theta)}{E^2}=\hat q^2-(\hat\lambda^2\csc^2\theta-a^2)\cos^2\theta.
\eea
\end{subequations}

Integrating up the geodesic equations for a photon trajectory from a source $(t_s, r_s, \theta_s, \phi_s)$ to an observer $(t_o, r_o, \theta_o, \phi_o)$, we obtain
\begin{subequations}
\label{eq:RayTracing}
\bea
\label{eq:RTheta}
&&-\kern-1.05em\int^{r_o}_{r_s}\frac{dr}{\pm\sqrt{\mathcal{R}(r)}}=
-\kern-1.05em\int^{\theta_o}_{\theta_s}\frac{d\theta}{\pm\sqrt{\Theta(\theta)}},\\
\label{eq:Phi}
\Delta\phi=\phi_o-\phi_s&=&-\kern-1.05em\int^{r_o}_{r_s}\frac{a}{\pm\Delta
\sqrt{\mathcal{R}(r)}}\Big(2Mr-a\hat{\lambda}\Big)dr\nn\\
&&+
-\kern-1.05em\int^{\theta_o}_{\theta_s}
\frac{\hat{\lambda}\csc^2\theta}{\pm\sqrt{\Theta(\theta)}}d\theta,\\
\label{eq:T}
\Delta t=t_o-t_s&=&-\kern-1.05em\int^{r_o}_{r_s}\frac{dr}{\pm\Delta
\sqrt{\mathcal{R}(r)}}\Big(\delta^2-a^2\Delta-2aMr\hat{\lambda}\Big)\nn\\
&&+-\kern-1.05em\int^{\theta_o}_{\theta_s}
\frac{a^2\cos^2\theta}{\pm\sqrt{\Theta(\theta)}}d\theta,
\eea
\end{subequations}
Here and hereafter, we use the subscript $o$ to denote the \lq\lq{}observer\rq\rq{}. We use the slash notation to denote that the integrals will be evaluated along a photon trajectory connecting the two points, since a turning point in the trajectory would occur every time
when the effective potential $\mathcal{R}(r)=0$ or $\Theta(\theta)=0$.
That is,
there might be more than one possibilities connecting the two points. Thus, we will introduce new parameters $n$, $m$ and $s$ to tell them apart which we will explain now. We use $n=0$ for those direct trajectories with no radial turning point and $n=1$ for those reflected trajectories with one radial turning point. As for the $\theta$ direction, we apply $m\geq0$ to denote the number of angular turning points and let $s=\pm1$ depict the final sign of $p_\theta$ (final vertical orientation). Then the trajectory equations can be reexpressed as the ``Kerr-Sen lens equations",
\begin{subequations}
\label{eq:LensEqns}
\bea
I_r+n\tilde{I}_r&=&G^{m,s}_{\theta},\label{first eq}\\ J_r+n\tilde{J}_r+\frac{\hat{\lambda}G^{m,s}_{\phi}-\Omega_sa^2G^{m,s}_t}
{M}&=&-\Omega_s t_o+2\pi N,\,\,\,\,\,\label{second eq}
\eea
\end{subequations}
where we have used $\phi_s=\Omega_s t_s$ to decouple $t_s$ from these equations and set $\phi_o=2\pi N$ for an integral $N$,
and $I_r,\ \tilde{I}_r,\ J_r,\ \tilde{J}_r$, and $G^{m,s}_i\ (i\in \{t,\theta,\phi\})$  are the radial and angular integrals given in App.~\ref{app:Integrals} which are defined in the same way as in Ref.~\cite{Gralla:2017ufe}.
\subsection{Observational appearance}
\label{subsec:ObservationalApp}
We will now consider the observational appearance of the point emitter following Refs.~\cite{cunningham1972optical,cunningham1973optical,Gralla:2017ufe}. The observables are the images positions, redshift factors and fluxes which can be expressed in terms of the conserved quantities \eqref{hatlambdaQ}.

The apparent position $(\alpha,\beta)$ of images on the observer's screen is obtained as
\begin{subequations}
\label{exact position}
\bea
\alpha&=&-\frac{\hat{\lambda}}{\sin\theta_o},\\
\beta&=&s\sqrt{\hat{q}^2+a^2\cos^2\theta_o-\hat{\lambda}^2\cot^2\theta_o}\nn\\
&&=s\sqrt{\Theta(\theta_o)},
\eea
\end{subequations}
where $s\in\{-1,1\}$ denotes the final sign of $p_\theta$ at the observer, which represents whether the photon arrives from above or below.

The ``redshift factor" $g$ is obtained as
\be
\label{redshift}
	g=\frac{E}{E_s}	=\frac{(\rho_b)_s}{\gamma}\sqrt{\frac{\Delta_s}{\Xi_s}}\frac{1}{1-\Omega_s\hat{\lambda}},
\ee
where the boost factor $\gamma$ is defined as
\bea
\label{vs}
v_s=\frac{\Xi_s}{\rho_b^2\sqrt{\Delta_s}}(\Omega_s-\omega_s),\,\,\,\,
\gamma=\frac{1}{\sqrt{1-v_s^2}}.
\eea

The normalized flux (comparing to the ``Newtonian flux" $F_N$) $F_o/F_N$ is given by
\be
\label{flux final pre}
\frac{F_o}{F_N}=\frac{g^3M(\rho_b)_{s}}{\gamma \sin\theta_o}\sqrt{\frac{\Delta_s}{\Xi_s\Theta(\theta_o)
\mathcal{R}(r_s)}}\Bigg|\det\frac{\partial(B,A)}{\partial(\hat{\lambda},
\hat{q})}\Bigg|^{-1},
\ee
where we have defined
\begin{subequations}
\label{eq:DefAB}
\bea
\label{eq:DefA}
A&\equiv& I_r+n\tilde{I}_r-G^{m,s}_{\theta}\pm M\int^{\theta_s}_{\pi/2}\frac{d\theta}{\sqrt{\Theta(\theta)}},\\
B&\equiv& J_r+n\tilde{J}_r+\frac{\hat{\lambda}G^{m,s}_{\phi}-\Omega_sa^2G^{m,s}_t}
{M}.
\eea
\end{subequations}
Note that computing the flux involves a variation regarding to $\theta_s$, thus we have generalized the lens equations \eqref{eq:LensEqns} to allow $\theta_s\neq\pi/2$.
The $\pm$ sign in Eq.~\eqref{eq:DefA} corresponds to pushing the source above/below the equatorial plane.

Like in the KM case \cite{Guo:2018kis}, these observables \eqref{exact position}, \eqref{redshift} and \eqref{flux final pre} also have same form as those of the Kerr case,
while the differences come from the specific expressions for $\Xi_s, (\rho_b)_s, \Delta_s$, $\omega_s$ and $\Omega_s$.

\section{Near-extremal expansion}
\label{sec:NearExtremalExpansion}
So far, we have set up the problem for the most general case. However it is not easy to analytically compute the emission signals for most of the situations, therefore, we will specialize to the case of an emitter orbiting on, or near, the prograde Innermost Stable Circular Orbit (ISCO) of a near-extremal KS black hole (as was suggested in Ref.~\cite{Gralla:2017ufe}).
Without loss of generality, we take the observer to sit in the northern hemisphere $\theta_o\in(0, \pi/2)$. 
For simplicity, we introduce a dimensionless radial coordinate $R$, which is related to the Boyer-Lindquist radius $r$ by
\be
R=\frac{r-(M-b)}{M}=\frac{r-M(1-\tilde{b})}{M},
\ee
where we have introduced the reduced twist parameter $\tilde{b}=b/M$. In addition, a small parameter $\epsilon$ is also introduced to describe the deviation of the KS black hole from extremity,
\be
\label{aN}
a=M(1-\tilde{b})\sqrt{1-\epsilon^3},
\ee

The ISCO of a general KS can be obtained by following the standard procedure given in Ref.~\cite{bardeen1972rotating}
and the leading order result in the near-extremal limit is obtained as (see App.~\ref{app:ISCO} for details)
\be
R_{\text{\tiny ISCO}}=\bar{R}\epsilon+\mathcal{O}(\epsilon^2),\label{rN}
\ee
where
\be
\bar{R}=2^{1/3}(1-\tilde{b})^{1/3}.\nonumber
\ee
Thus, to the leading order in $\epsilon$, the source orbits on the radius
\be
\label{eq:RsExpansion}
r_s=M(1-\tilde{b}+\epsilon \bar{R}).
\ee

In order to keep track of the small corrections, we introduce a new quantity $\lambda$ instead of $\hat{\lambda}$ in a convenient manner \cite{Gralla:2017ufe}, as
\bea
\hat{\lambda}&=&2M(1-\epsilon\lambda).
\eea
Following Ref.~\cite{porfyriadis2017photon}, a new quantity $q$ is also introduced as
\be
\hat{q}=M\sqrt{(1-\tilde{b})(3+\tilde{b})-q^2}.
\ee

For later reference, we now expand the orbital frequency $\Omega_s$ and period $T_s$ in $\epsilon$, which are given by
\be
\Omega_s=\frac{1}{2M}+\mathcal{O}(\epsilon),\qquad
T_s=\frac{2\pi}{\Omega_s}=4\pi M+\mathcal{O}(\epsilon).
\ee
To leading order in the deviation of near-extremality, the orbital frequency remains unchanged for the KS case compared to the Kerr case, which is different from the KM \cite{Guo:2018kis} and KN (App.~\ref{app:KN}) cases.

\subsection{Near-extremal solutions}
\label{subsec:Solution}
Now we solve the Kerr-Sen lens equations \eqref{eq:LensEqns} in the near-extremal limit and write the solutions $(\lambda,q)$ as functions of observer's time $t_o$.

\subsubsection{First equation}
We start with the first equation \eqref{first eq},
\be
\label{final first eq}
I_r+n\tilde{I}_r=mG_{\theta}-s\hat{G_\theta}.
\ee
The $I$ integrals (radial) and $G$ integrals (angular) are performed in App.~\ref{app:Integrals}.
The results for the radial integrals are given by
\begin{subequations}
\label{eq:Iintegral}
\bea
I_r&=&\frac{1}{q}
\log\Bigg[\frac{4q^4R_o}{(qD_o+q^2+2R_o)(qD_s+q^2
\bar{R}+4(1-\tilde{b})\lambda)}\Bigg]\nn\\
&&-\frac{1}{q}\log\epsilon+\mathcal{O}(\epsilon),\\
\tilde{I}_r&=&\frac{1}{q}\log\Bigg[\frac{(qD_s+q^2
\bar{R}+4(1-\tilde{b})\lambda)^2}{4(1-\tilde{b})^2(4-q^2)\lambda^2}
\Bigg]+\mathcal{O}(\epsilon),
\eea
\end{subequations}
where
\bea\label{eq:Ds}
D_s&=&\sqrt{q^2\bar{R}^2+8(1-\tilde{b})\lambda\bar{R}+4(1-\tilde{b})^2\lambda^2},\\
\label{eq:Do}
D_o&=&\sqrt{q^2+4R_o+R_o^2}.
\eea
The results for the angular integrals are given by elliptic functions which are some explicit functions of $q$.

We will introduce the main steps for solving this equation and refrain from giving detailed calculation since a similar calculation can be found in Ref.~\cite{Gralla:2017ufe}.

First, we introduce two quantities $\bar{m}$ and $\Upsilon >0$ for convenience, which are defined by
\bea
\label{eq:mbar}
m&=&-\frac{1}{qG_{\theta}}\log\epsilon+\bar{m},\\
\label{eq:Upsilon}
\Upsilon&\equiv&\frac{q^4R_o}{q^2+2R_o+qD_o}e^{-qG^{\bar{m},s}_{\theta}}.
\eea
The logarithmic term in \eqref{eq:mbar} is introduced to compensate for the corresponding term in LHS of Eq.~\eqref{final first eq}.

Next, we can rewrite the equation \eqref{final first eq} in a simplified form by using \eqref{eq:mbar} and \eqref{eq:Upsilon}, which leads to a quadratic equation in $\lambda$.
Solving this quadratic equation for $n=0$ and $n=1$, respectively, we obtain the final result as
\be
\label{lambda of q}
\lambda=\frac{2\Upsilon}{(1-\tilde{b})(4-q^2)}\Bigg[2- q\sqrt{1+\frac{4-q^2}{2\Upsilon}\bar{R}}\Bigg].
\ee
In addition, the
conditions for the solution to exist are obtained as
\begin{subequations}
\label{conditions}
\bea
\bar{R}&<&\frac{4\Upsilon}{q^2}\Bigg(1+\frac{2}{\sqrt{4-q^2}}\Bigg)
\quad \text{if}\quad n=0 ,\\
\bar{R}&>&\frac{4\Upsilon}{q^2}\Bigg(1+\frac{2}{\sqrt{4-q^2}}\Bigg)
\quad \text{if}\quad n=1.
\eea
\end{subequations}

We discover that the first equation \eqref{final first eq} does not include in explicity the time $t_o$. Thus, for given choice of $m$, $s$, and $n$, we have arrived at a function $\lambda(q)$ [Eq.~\eqref{lambda of q}] with supplementary conditions \eqref{conditions}.

\subsubsection{Second equation}
From the second equation \eqref{second eq} we will get another
another relation between $t_o$, $\lambda$, and $q$ for given choice of $m$, $s$, and $n$. For convenience, we introduce a dimensionless time coordinate $\hat{t}_o$ restricted to a single period $\hat{t}_o\in[0,1]$,
\be
\hat{t}_o=\frac{t_o}{T_s}=\frac{t_o}{4\pi M}+\mathcal{O}(\epsilon).
\ee
We can then rewrite Eq.~\eqref{second eq} as
\be
\label{definition G}
\hat{t}_o=N+\mathcal{G},\,\,
\mathcal{G}=-\frac{1}{2\pi}\Big(J_r+n\tilde{J}_r+2G^{m,s}_{\phi}
-\frac{(1-\tilde{b})^2}{2}G^{m,s}_{t}\Big),
\ee
where $J$ integrals and $G$ integrals are given in App.~\ref{app:Integrals}. The $J$ integrals have similar structures as the $I$ integrals \eqref{eq:Iintegral} and $G$ integrals are functions of $q$. Recalling Eq.~\eqref{lambda of q}, we can then conclude that Eq.~\eqref{definition G} gives a function
\be
\label{eq:tofq}
\hat{t}_o(q)=\hat{t}_o[q,\lambda(q)]
\ee
for a given choice of $m,\,s,\,b$ with a non-zero range of q. Note also that in the given period of $\hat{t}_o\in[0,1]$, $N$ is uniquely determined.

For each allowed value of $N$, inverting Eq.~\eqref{eq:tofq} in each monotonic domain gives multivalued inverses $q_i(\hat{t}_o)$ with $i$ a discrete integral labeling each of these inverses. We know from Sec.~\ref{subsec:ObservationalApp} that the observables of a hot spot's image can be written as functions of $\lambda$ and $q$, therefore, each function $q(\hat{t}_o)$ corresponds to an image track labeled by $(m,\,n,\,s,\,N,\,i)$.

\subsection{Observational quantities}
\label{subsec:NearExtremalApp}
In this subsection, we will expand the observables, including the position \eqref{exact position}, the redshift factor \eqref{redshift} and the flux \eqref{flux final pre}, in $\epsilon$ and pick up the leading piece. Recall from the beginning of Sec. \ref{sec:NearExtremalExpansion}, we have the near-extremal KS expansions
\bea
\label{eq:RecallExpansion}
a=M(1-\tilde{b})\sqrt{1-\epsilon^3},\quad
r_s=M(1-\tilde{b}+\epsilon \bar{R}),\nn\\
\hat{\lambda}=2M(1-\epsilon\lambda),\quad
\hat{q}=M\sqrt{(1-\tilde{b})(3+\tilde{b})-q^2}.
\eea

By expanding Eq.~\eqref{exact position}, the image position (impact parameters) on the observer's screen is obtained as
\begin{subequations}
\label{image position}
\bea
\alpha&=&-\frac{2M}{\sin\theta_o}+\mathcal{O}(\epsilon),\\
\beta&=&sM\Big[3-q^2+\cos^2\theta_o-4\cot^2\theta_o\nn\\
&&-\tilde{b}
(3+\cos2\theta_o+\tilde{b}\sin^2\theta_o)\Big]^{1/2}+\mathcal{O}(\epsilon).
\eea
\end{subequations}
Note that
$\lambda$ does not appear in the position and the impact parameter $\alpha$ is the same as in the Kerr case.
The other impact parameter $\beta$ being real gives a range of $q$,
\be
\label{range of q}
q\in\Big[0,\sqrt{3+\cos^2\theta_o-4\cot^2\theta_o-\tilde{b}
(3+\cos2\theta_o+\tilde{b}\sin^2\theta_o)}\Big].\nn
\ee
The boundary values of $q$ corresponds to two endpoints of a vertical line (the analogue of NHEK line \cite{Gralla:2017ufe}), out of which the image disappears since $\beta$ is no longer real. For an observer being able to see the image, the inclination is required to be in the range of $\theta_{\text{crit}}<\theta_o<\pi-\theta_{\text{crit}}$, where
\be
\label{eq:criticalAngle}
\theta_{\text{crit}}=\arctan\Bigg[\frac{2}{\sqrt{2\sqrt{3-2\tilde{b}}-2\tilde{b}}}
\Bigg].
\ee

Next, the redshift factor \eqref{redshift} is expanded as
\bea
\label{redcorr}
g=\frac{\sqrt{(1+\tilde{b})(3+\tilde{b})}}{(3+\tilde{b})+4\frac{\lambda}{\bar{R}}}
+\mathcal{O}(\epsilon).
\eea

The normalized image flux \eqref{flux final pre} is expanded as,
\bea
\label{flux extremal}
\frac{F_o}{F_N}&=&\frac{\sqrt{1+\tilde{b}}\bar{R}\epsilon}{2\sqrt{1-\tilde{b}}D_s}
\frac{qg^3}{\sin\theta_o\sqrt{1-\frac{q^2}{3-2\tilde{b}-\tilde{b}^2}}
\sqrt{\Theta_0(\theta_o)}}\nn\\
&&\times\Bigg|\det\frac{\partial(B,A)}
{\partial(\lambda,q)}\Bigg|^{-1},
\eea
where $g$ and $D_s$ are given in Eqs.~\eqref{redcorr} and \eqref{eq:Ds}, and (see Eq.~\eqref{image position})
\bea
&&\Theta_0(\theta_o)=[\Theta(\theta_o)/M^2]|_{\lambda=0}=\beta^2/M^2\nn\\
&=&3-q^2+\cos^2\theta_o-4\cot^2\theta_o-\tilde{b}
(3+\cos2\theta_o+\tilde{b}\sin^2\theta_o),\,\,\,\,\,\,\,\,\,\,
\eea
and from the definitions of $A$ and $B$ [Eq.~\eqref{eq:DefAB}], we have
\bea
\label{eq:JacabiBLambda}
\left|\det\frac{\partial (B,A)}{\partial (\lambda,q)}\right|&=&
\bigg|\frac{\partial}{\partial \lambda}\bigg(J_r+b\tilde{J}_r\bigg)\bigg[\frac{\partial}{\partial q}\bigg(I_r+b\tilde{I}_r\bigg)-\frac{\partial G^{m,s}_{\theta}}{\partial q}\bigg]\nn\\
&&-\frac{\partial}{\partial\lambda}\bigg(I_r+b\tilde{I}_r\bigg)\bigg[
\frac{\partial}{\partial q}\bigg(J_r+b\tilde{J}_r\bigg)+\frac{\partial G^{m,s}_{t\phi}}{\partial q}\bigg]\bigg|,\nn\\
&&+\mathcal{O}(\epsilon\log\epsilon),
\eea
where we have introduced
\bea
G^{m,s}_{t\phi}&=&\big(\hat{\lambda}G^{m,s}_{\phi}-\Omega_s a^2G^{m,s}_t\big)/M\nn\\
&=&2G^{m,s}_{\phi}-\frac{(1-\tilde{b})^2}{2}G^{m,s}_t+\mathcal{O}(\epsilon).
\eea
The $G$, $I$, $J$ integrals are given in App.~\ref{app:Integrals}.

\section{Results and discussion}
\label{sec:ResultsDiscussion}
To make the results of the observables more practical, we will make the following choices for the parameters,
\bea
\label{eq:Parameters}
R_o&=&100,\qquad \theta_o=84.27^{\circ},\nn\\
\epsilon&=&0.01,\qquad \bar{R}=\bar{R}_{\text{ISCO}}=2^{1/3}\big(1-\tilde{b}^2\big)^{1/3}.
\eea
In addition, for the reduced twist parameter $\tilde{b}$, we will restrict ourself to a range from $0$ to $0.5$. As mentioned before, the $\tilde{b}=0$ case reduce to the Kerr one. While we take the upper bound $\tilde{b}=0.5$ corresponding to the $U(1)$ charge $Q=M$, since the charge for an astrophysical black hole is supposed to be quite small. (Note also, from App.~\ref{app:KN}, that the Kerr-Newman metric with a non-zero spin and a $U(1)$ charge $Q_{\text{KN}}=M_{\text{KN}}$ represents a naked singularity whose apparent shape has been studied in Ref.~\cite{PhysRevD.78.044007}.) Thus, we have set up a hot spot on the ISCO of a near-extremal KS black hole with a charge $Q$ ranging from $0$ to $M$, viewed from a nearly edge-on inclination by a distant observer. Here we have used subscript `KN' for physical charges of KN spacetime and we will use subscipt `KM' for those of KM spacetime. Charges without subscript are for KS spacetime.

Moreover, we will compare the results for KS black hole with those for other charged rotating black holes (the KN and KM black holes). The result for KM case can be found in Ref.~\cite{Guo:2018kis} and the result for KN case is introduced in App.~\ref{app:KN}.

The main observables of the hot spot image in near-extremal KS spacetime are given in Sec.~\ref{subsec:ObservationalApp}, which are the apparent position $(\alpha,\beta)$ \eqref{image position}, the redshift factor $g$ \eqref{redcorr} and the flux $F_o/F_N$ \eqref{flux extremal}. The complete information of the image consists of pieces from the image segments labeled by $(m,n,s,N,i)$ for all choices of these parameters (see Sec.~\ref{sec:NearExtremalExpansion} for details). Next, we will illustrate the feature of these observables for several selected brightest images.
\begin{figure*}[ht!]
\begin{center}
\includegraphics[width=\textwidth]{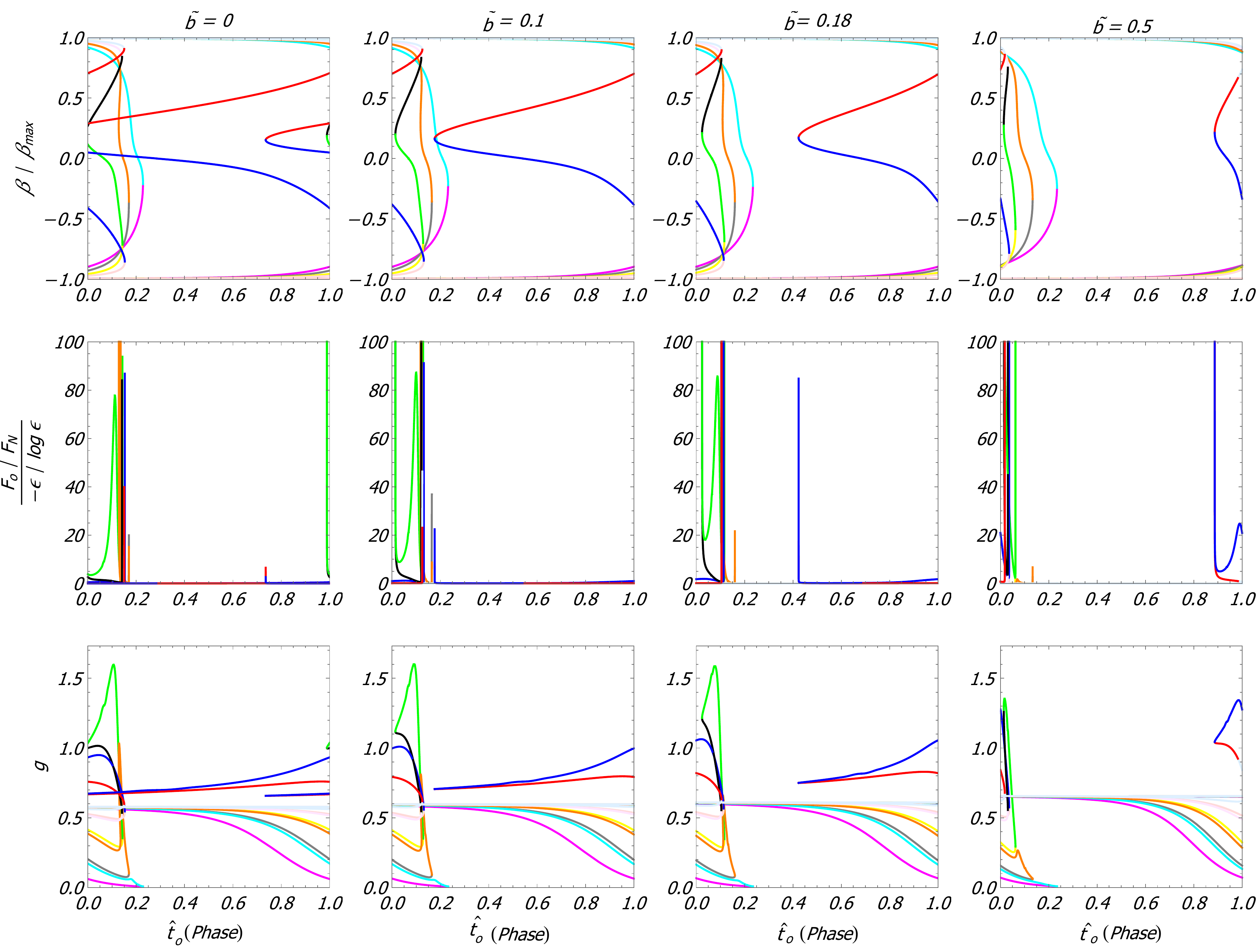}
\end{center}
  \caption{Positions, fluxes and redshift factors of the few brightest images for the twist parameter $\tilde{b}=0$ (Kerr case), 0.1, 0.18 ($Q=0.6M$) and 0.5 ($Q=M$). We have color-coded by continuous image tracks labeled by $(m,\,n,\,s,\,N,\,i)$ following Ref.~\cite{Gralla:2017ufe}.}
  \label{fig:together}
\end{figure*}

In Fig.~\ref{fig:together}, we show the main observables of these selected bright images in a single period for the reduced twist parameter $\tilde{b}=0$, $0.1$, $0.18$, and $0.5$, respectively. Note that we choose the specific value of $\tilde{b}=0.18$ because it corresponds to the Sen charge $Q=0.6M$, which will be compared, as a representative example, with the KN case for the KN charge $Q_{\text{KN}}=0.6M_{\text{KN}}$.
For each $\tilde{b}$, each graph has several different colored lines of which the green one represents the primary image while the others indicate the secondary images. The secondary images are much fainter than the primary image apart from near caustics (where different line segments intersect). The general features of the primary image (which moves on a vertical line while blueshifting and spiking in brightness) and the secondary images (which are also on the vertical line with a rich caustic structure) are the same as those for the Kerr case \cite{Gralla:2017ufe}. Note that the graphs for $\tilde{b}=0$ agree exactly with \cite{Gralla:2017ufe} which is no surprise since the KS metric reduce to the Kerr metric in that case. However, when $\tilde{b}\neq0$, the KS case displays quantitatively corrections to the kerr case and also has differences from the KM (KN) case \cite{Guo:2018kis}. Regarding to the apparent position \eqref{image position}, the impact parameter $\alpha$ stays unchanged when $\tilde{b}$ is varied, while the maximum value of the other impact parameter $\beta$ decreases when $\tilde{b}$ is increased from $0$ (corresponding to $\beta_{\text{max}}=1.72M$) to $0.5$ (corresponding to $\beta_{\text{max}}=1.31M$). From the middle line we see that the energy flux increases when $\tilde{b}$ is increased. From the bottom line we see that, when $\tilde{b}$ is increased, the peak redshift factor associated with the primary image stays around $1.6$ (except for $\tilde{b}=0.5$) while the typical redshift factor (corresponding to $\lambda\sim0$) associated with the secondary images increases. To be specific, one can obtain this typical redshift factor from Eq.~\eqref{redcorr}, as $g=\sqrt{(1+\tilde{b})/(3+\tilde{b})}$. The astronomical observed iron line \cite{carlson20153} might suppose to be shifted by these factors. Unfortunately, comparing with the predicted value in the Kerr case \cite{Gralla:2017ufe}, the results obtained in the KS cases are further away from the observed value. These observational signatures appear periodically and the period in KS cases stay unchanged when $\tilde{b}$ is varied. In addition, as in the Kerr case, these signatures are strongest in the edge-on case ($\theta_o\approx90^{\circ}$) and will disappear when the inclination $\theta_o$ less than a critical angle $\theta_{\text{crit}}$ (see Eq.~\eqref{eq:criticalAngle}). In the KS cases, this inclination angle is increased from $47^{\circ}$ to $56^{\circ}$ when $\tilde{b}$ is increased from $0$ to $0.5$.

Furthermore, now we illustrate the representative example for the observational signatures of near-extremal KS/KN black holes both with the same $U(1)$ charge $Q_{\text{(KN)}}=0.6M_{\text{(KN)}}$. As mentioned before, this corresponds to the reduced twist parameter $\tilde{b}=0.18$ for the KS case. The results for the KN case can be obtained from a comparison with the KM case \cite{Guo:2018kis} which has a mass-dependent charge (see App.~\ref{app:KN}). We find that, in KS and KN cases, the charges $Q_{\text{(KN)}}$ trend to have positively correlated influences on most of the observables and these observables are corrected more in the KN case. However, the impact parameter $\alpha$ and the period $T_s$ in the KS case stay unchanged but they both are corrected in the KN case. We show these results explicitly in Table~
\ref{table:KNvsKS}.
\begin{table}
\centering
\begin{tabular}{|l|l|l|l|}
  \hline
  &Kerr & KS & KN \\
  \hline
  $Q$& 0& 0.6M& 0.6M\\
  $\alpha$&$-20M$& $-20M$ & $-20.53M$ \\
  $\beta_{\text{max}}$& $1.72M$ & $1.60M$ & $1.55M$ \\
  $g_{\text{peak}}$ &1.6& 1.6 & 1.6 \\
  $g_{\lambda\sim0}$ & 0.58 & 0.61 & 0.64 \\
  $T_s$ & $4\pi M$ &$4\pi M$ & $4.1\pi M$ \\
  $\theta_{\text{crit}}$ & $47.06^{\circ}$ & $49.66^{\circ}$ & $51.35^{\circ}$ \\
  \hline
\end{tabular}
  \caption{Some typical quantities for the near-extremal KS and KN black hole both with a $U(1)$ charge $Q=0.6M$ ($Q_{\text{KN}}=0.6M_{\text{KN}}$) and the corresponding quantities for the neutral Kerr case.}
  \label{table:KNvsKS}
\end{table}

\subsection*{Acknowledgements}
The authors would like to thank Niels A. Obers for helpful comments and for reading the manuscript.
MG is partially supported by NSFC with Grants No. 11675015, No. 11775022, No. 11875095 and No. 11947210. He is also funded by China National Postdoctoral Innovation Program 2019M660278. HY thanks the Theoretical Particle Physics and Cosmology section at the Niels Bohr Institute for support. SS and HY are also financially supported by the China Scholarship Council.

\appendix
\section{Kerr-Newman black hole revisited and its observational signature}
\label{app:KN}
A useful reference for understanding the charged black hole in the string theory [the Kerr-Sen (KS) black hole] is to study its counter partner in the general relativity [the Kerr-Newman (KN) black hole].
Therefore, here we briefly review the KN metric and revisit it from a mathematical comparison with the Kerr-MOG (KM) metric. We do the later comparison because the observational signature of a near-extremal KM black hole have been studied in Ref.~\cite{Guo:2018kis} and we will see from below that the expressions for the near-extremal KM case can be applied to the near-extremal KN case upon replacements of the corresponding parameters in these metrics.

The KN metric is a stationary solution of the Einstein-Maxwell theory, which in Boyer-Lindquist coordinates reads
\bea
\label{eq:KNmetric}
ds^2&=&-\frac{\Delta_{\text{KN}}}{\Sigma_{\text{KN}}}\big(dt-\tilde{a}\sin^2\theta d\phi\big)^2+\frac{\Sigma_{\text{KN}}}{\Delta_{\text{KN}}}dr^2\nn\\
&&+\Sigma_{\text{KN}}
d\theta^2+\frac{\sin^2\theta}{\Sigma_{\text{KN}}}\big(\tilde{a}dt-(r^2
+\tilde{a}^2)d\phi\big)^2,
\eea
where
\be
\Sigma_{\text{KN}}=r^2+\tilde{a}^2\cos^2\theta,\,\,
\Delta_{\text{KN}}=r^2-2M_{\text{KN}}r+\tilde{a}^2+Q_{\text{KN}}^2,
\ee
with $\tilde{a}$, $M_{\text{KN}}$, and $Q_{\text{KN}}$ being the spin, mass, and $U(1)$ charge of the black hole. For later reference, we may define a reduced charge parameter $\tilde{e}=Q_{\text{KN}}/M_{\text{KN}}$. When $Q_{\text{KN}}=0$, this reduce to the Kerr black hole.
The extremal limit for KN black hole is obtained for
\be
\label{eq:ExtreKN}
\tilde{a}^2=M_{\text{KN}}^2-Q_{\text{KN}}^2.
\ee

For a comparison of the KS metric and KN metric from the action level and a comparison of their apparent shapes, readers may refer to Ref.~\cite{PhysRevD.78.044007}.

Next, we introduce the KM metric which is a stationary solution of
the scalar-tensor-vector (STVG) modified gravitational (MOG) theory. We will restore the Newtonian constant $G_N$ to describe the KM metric for the reason that will become clear below. The KM metric in Boyer-Lindquist coordinates reads
\bea
\label{eq:KMmetric}
ds^2&=&-\frac{\Delta_{\text{KM}}}{\Sigma_{\text{KM}}}\big(dt-a\sin^2\theta d\phi\big)^2+\frac{\Sigma_{\text{KM}}}{\Delta_{\text{KM}}}dr^2\nn\\
&&+\Sigma_{\text{KM}} d\theta^2+\frac{\sin^2\theta}{\Sigma_{\text{KM}}}\big(adt-(r^2+a^2)d\phi\big)^2,
\eea
where
\be
\Sigma_{\text{KM}}=r^2+a^2\cos^2\theta,\,\,
\Delta_{\text{KM}}=r^2-2M_{\alpha}r+a^2+\beta^2,
\ee
with
\be
\label{beta}
M_{\alpha}=G_N(1+\alpha)m,\qquad
\beta^2=\frac{\alpha}{1+\alpha}M_{\alpha}^2.
\ee
Here, $m$ and $a$ are mass and spin parameters of the
black hole and $\alpha$ is the deformation parameter defined by $G=G_N (1+\alpha)$ with $G$ being an enhanced gravitational constant. Moreover, $M_{\alpha}$ and $J=M_{\alpha}a$ are, respectively, the ADM mass and angular momentum of the KM metric \cite{sheoran2017mass}. In addition, $K=\sqrt{\alpha  G_N}m$ is defined as the gravitational charge of the MOG vector field and $\beta$ in \eqref{beta} is a parameter related to this charge. Note that, unlike the $U(1)$ charges of the KS and KN black holes which are independent from their masses, this gravitational charge in the MOG theory is mass-dependent.
The extremal limit for KM black hole is obtained for
\be
\label{eq:ExtreKM}
a^2=M_{\alpha}^2-\beta^2.
\ee

As briefly introduced above, the physical starting points of the KN and KM metrics and their interpretations are both quite different. Nevertheless, it is interesting that the KN metric \eqref{eq:KNmetric} and the KM metric \eqref{eq:KMmetric} have very similar mathematical forms of their expressions. We will interpret $M_\alpha$ as the mass of the KM black hole as suggested in Ref.~\cite{sheoran2017mass} corresponding to $M_{\text{KN}}$ in the KN black hole, then the only difference between these metrics comes from the mass-dependencies of their charge parameters. Furthermore, this difference become irrelevant in the expressions for the (near-)extremal cases due to the constraints \eqref{eq:ExtreKN} and \eqref{eq:ExtreKM}. In those cases, we may mathematically identify the KN metric and KM metric upon $M_{\text{KN}}\rightarrow M_\alpha$ and $Q_{\text{KN}}\rightarrow\beta(=\sqrt{M_{\alpha}^2-a^2})$.
As have been shown in the main text (as well as in Refs.~\cite{Gralla:2017ufe,Guo:2018kis}) that the computations for the observables of an orbiting emitter on the ISCO of a near-extremal rotating black hole only rely on the spacetime metric and the geodesics in it. Therefore, the results in Ref.~\cite{Guo:2018kis} for the KM case can be applied to the KN case upon to the replacements: $M_\alpha\rightarrow M_{\text{KN}}$ and $\sqrt{M_{\alpha}^2-a^2}\rightarrow Q_{\text{KN}}$. Note that in Ref.~\cite{Guo:2018kis} it has set $M_\alpha=1$ and used spin $a$ to represent the modified parameter $\alpha$ due to the relations \eqref{beta} and \eqref{eq:ExtreKM}. To be specific, the near-extremal KM cases for $a=1$, $0.8$ and $0.717$ correspond to the near-extremal KN cases for $\tilde{e}=0$, $0.6$ and $0.7$. As a particular and representative example to be compared with the corresponding KS case, we have discussed the KN case for $\tilde{e}=0.6$ in Sec.~\ref{sec:ResultsDiscussion}, for which we have borrowed the results of the KM one for $a=0.8$ from Ref.~\cite{Guo:2018kis}.

\section{ISCO of a near-extremal Kerr-Sen black hole}
\label{app:ISCO}
The innermost stable circular orbit of a particle in the equatorial plane can be obtained by instantaneously solving the following equations \cite{bardeen1972rotating},
\be
\label{eq:appISCO}
\tilde{\mathcal{R}}(r)=0, \qquad
\frac{\mathrm{d}\tilde{\mathcal{R}}(r)}{\mathrm{d}r}=0,\qquad
\frac{\mathrm{d^2}\tilde{\mathcal{R}}(r)}{\mathrm{d}r^2}=0,
\ee
where $\tilde{\mathcal{R}}(r)$ is given in \eqref{eq:radialPoten}.

For the near-extremal Kerr-Sen case, we have
\be
\label{eq:appaN}
a=M(1-\tilde{b})\sqrt{1-\epsilon^3}.
\ee
In order to solve the radius of ISCO in this case to the $\mathcal{O}(\epsilon)$ order and note that the result should reduce to that in the near-extremal Kerr case when $\tilde{b}=0$ \cite{Gralla:2017ufe}, we then assume $r_{\text{ISCO}}$ has a form of
\be
\label{eq:iscoAssum}
r_{\text{ISCO}}=M(1-\tilde{b})+M f(M, \tilde{b})\epsilon
\ee
with $f(M,\tilde{b})$ a undetermined function. Solving Eqs.~\eqref{eq:appISCO} under the condition \eqref{eq:appaN} and the assumption \eqref{eq:iscoAssum}, we obtain
\bea
f(M, \tilde{b})=2^{1/3}(1-\tilde{b})^{1/3}.
\eea

\section{Integrals}
\label{app:Integrals}
\subsection{Radial integrals and matched asymptotic expansion method}
\label{app:MAEandRadialIntegral}
The radial integrals appearing in the ``Kerr-Sen lens equations" \eqref{eq:LensEqns} are defined in the same way as \cite{Gralla:2017ufe}
\begin{subequations}
\label{eq:RadialIntegrals}
\bea
\label{eq:IntergralIr}
I_r&=&M\int^{r_o}_{r_s}\frac{dr}{\sqrt{\mathcal{R}(r)}},\,
\tilde{I}_r=2M\int^{r_s}_{r_{\text{min}}}\frac{dr}{\sqrt{\mathcal{R}(r)}},\\
\label{eq:IntergralJr}
J_r&=&\int^{r_o}_{r_s}\frac{\mathcal{J}_r}{\sqrt{\mathcal{R}(r)}}dr,\,\,\,\,
\tilde{J}_r=2\int^{r_s}_{r_{\text{min}}}\frac{\mathcal{J}_r}{\sqrt{\mathcal{R}(r)}}dr,
\eea
\bea
\label{eq:MathcalJr}
\mathcal{J}_r=\frac{1}{\Delta}\Big[a(2Mr-a\hat{\lambda})-\Omega_s
\big(\delta^2-a^2\Delta-2aMr\hat{\lambda}\big) \Big],\,\,\,\,
\eea
\end{subequations}
where $\delta$ and $\Delta$ are defined in \eqref{eq:MetricPara}, $\Omega_s$ is defined in \eqref{eq:AngularVelocity}, $\mathcal{R}(r)$ is defined in \eqref{eq:NullRadialPoten} and $r_{\text{min}}$ is the largest (real) root of $\mathcal{R}(r)=0$. These equations are valid when $r_{\text{min}}<r_s$, which is always true for a light that can reach infinity.

These radial integrals \eqref{eq:RadialIntegrals} can be approximately performed, to the leading order in $\epsilon$ in the near-extremal limit, by using the matched asymptotic expansion (MAE) method which was introduced in Refs.~\cite{porfyriadis2017photon,Gralla:2017ufe} (this MAE method has also been applied to the large dimension limit in Ref.~\cite{Guo:2019photon} recently).

Next, we will take $I_r$ for example to show the technical steps.
For convenience and self-consistence in this appendix, we cope the relevant formulas to here:
\bea
\mathcal{R}(r)&=&(a\hat\lambda-\delta)^2-\Delta\left[\hat q^2+(\hat\lambda-a)^2\right],\nn\\
\delta&=&r^2+a^2+2\tilde{b}Mr,\nn\\
\Delta&=&\delta-2Mr.
\eea
We will use the dimensionless radial coordinate, $R=(r-M+M\tilde{b})/M$, for convenience. In the near-extremal limit, we have the expansions (from the beginning of Sec.~\ref{sec:NearExtremalExpansion}),
\bea
a=M(1-\tilde{b})\sqrt{1-\epsilon^3},\quad
r_s=M(1-\tilde{b}+\epsilon \bar{R}),\nn\\
\hat{\lambda}=2M(1-\epsilon\lambda),\quad
\hat{q}=M\sqrt{(1-\tilde{b})(3+\tilde{b})-q^2}.
\eea
By introducing two constants $0<p<1$ and $C>0$, we split the $I_r$ integral into two pieces,
\be
I_r=M^2\int^{\epsilon^p C}_{\epsilon\bar{R}}\frac{dR}{\sqrt{\mathcal{R}}}+M^2\int^{R_o}_{\epsilon^p C}\frac{dR}{\sqrt{\mathcal{R}}}.
\ee
The scaling of $\epsilon^p$ introduces a separation of scales $\epsilon\ll\epsilon^p\ll1$ as $\epsilon\rightarrow0$, such that the first piece of integral is in the near horizon region $R\sim\epsilon$ and the second piece is in the far region $R\sim1$.

In the near horizon region, we make the change of variables $x=R/\epsilon$ and expand in $\epsilon$ at fixed $x$. Thus the first piece of integral is:
\bea
\label{eq:NearRegion}
&&M^2\int^{\epsilon^p C}_{\epsilon\bar{R}}\frac{dR}{\sqrt{\mathcal{R}}}\nn\\
&=&
M^2\int^{\epsilon^{p-1}C}_{\bar{R}}\Big[\frac{dx}{\sqrt{q^2x^2+8(1-\tilde{b})\lambda x+4(1-\tilde{b})^2\lambda^2}}+\mathcal{O}(\epsilon)\Big]\nn\\
&=&\frac{1}{q}\log\Big[\frac{2q^2}{qD_s+q^2
\bar{R}+4(1-\tilde{b})\lambda}+(p-1)\log\epsilon+\log C\Big]\nn\\
&&+\mathcal{O}(\epsilon^p),
\eea
where 
\be
\label{eq:appDs}
D_s=\sqrt{q^2\bar{R}^2+8(1-\tilde{b})\lambda\bar{R}+4(1-\tilde{b})^2\lambda^2},
\ee

In the far region, we expand in $\epsilon$ at fixed $R$. The second integral then reads:
\bea
\label{eq:FarRegion}
M^2\int_{\epsilon^p C}^{R_o}\frac{dR}{\sqrt{\mathcal{R}}}
&=&M^2\int_{\epsilon^p C}^{R_o}\Big[\frac{dR}{R\sqrt{q^2+4R+R^2}}+\mathcal{O}(\epsilon)\Big]\nn\\
&=&\frac{1}{q}\log\Big[\frac{2q^2R_o}{qD_o+q^2+2R_o}-p\log\epsilon-\log C\Big]\nn\\
&&+\mathcal{O}(\epsilon^p),
\eea
where
\be
\label{eq:appDo}
D_o=\sqrt{q^2+4R_o+R_o^2}.
\ee

By adding up Eqs. \eqref{eq:NearRegion} and \eqref{eq:FarRegion}, we get the complete integral:
\bea
I_r&=&\frac{1}{q}
\log\Big[\frac{4q^4R_o}{(qD_o+q^2+2R_o)(qD_s+q^2
\bar{R}+4(1-\tilde{b})\lambda)}\Big]\nn\\
&&-\frac{1}{q}\log\epsilon+\mathcal{O}(\epsilon),
\eea
where $D_s$ and $D_o$ are given in \eqref{eq:appDs} and \eqref{eq:appDo}.
Note that the constants $p$ and $C$ are cancelled in the final result.

The remaining radial integrals in \eqref{eq:RadialIntegrals} can be obtained in a similar way using the MAE method.
We now list all of the results for these integrals which appear in the Kerr-Sen lens equations \eqref{eq:LensEqns}, as follows:
\begin{subequations}
\bea
I_r&=&\frac{1}{q}
\log\Big[\frac{4q^4R_o}{(qD_o+q^2+2R_o)(qD_s+q^2
\bar{R}+4(1-\tilde{b})\lambda)}\Big]\nn\\
&&-\frac{1}{q}\log\epsilon+\mathcal{O}(\epsilon),\\
\tilde{I}_r&=&\frac{1}{q}\log\Big[\frac{(qD_s+q^2
\bar{R}+4(1-\tilde{b})\lambda)^2}{4(1-\tilde{b})^2(4-q^2)\lambda^2}
\Big]+\mathcal{O}(\epsilon),\\
\label{eq:IntegralJr}
J_r&=&-\frac{7-2\tilde{b}-\tilde{b}^2}{2}I_r-\frac{1}{2}(D_o-q)-\frac{3+\tilde{b}}{8}
\Big(\frac{q\bar{R}}{\lambda}-\frac{D_s}{\lambda}\Big)\nn\\
&&+\log\Big[\frac{(q+2)^2
\bar{R}}{(D_o+R_o+2)(D_s+2\bar{R}+2(1-\tilde{b})\lambda)}\Big]\nn\\
&&+\mathcal{O}(\epsilon),\\
\label{eq:IntegralJrTilde}
\tilde{J}_r&=&-\frac{7-2\tilde{b}-\tilde{b}^2}{2}\tilde{I}_r-\frac{3+\tilde{b}}{4}
\frac{D_s}{\lambda}\nn\\
&&+\log\Big[
\frac{(D_s+2\bar{R}+2(1-\tilde{b})\lambda)^2}{(4-q^2)\bar{R}^2}\Big]
+\mathcal{O}(\epsilon),
\eea
\end{subequations}
where $D_s$ and $D_o$ are given in \eqref{eq:appDs} and \eqref{eq:appDo}.
If we take $\tilde{b}=0$, the above results are exactly the same as those of Kerr \cite{Gralla:2017ufe}.
\subsection{Angular integrals}
\label{app:AngularIntegrals}
We are also interested in the angular integrals appearing in the lens equations~\eqref{eq:LensEqns} defined in the same way as Ref.~\cite{Gralla:2017ufe}
\begin{align}
\label{eq:AngleIntegralMS}
	G^{m,s}_i=
	\begin{cases}
		\hat{G}_i\qquad\qquad&m=0,\\
		mG_i-s\hat{G}_i\qquad&m\ge1,
	\end{cases}
	\qquad
	i\in\cu{t,\theta,\phi}.
\end{align}
with
\be
G_i=M\int^{\theta_+}_{\theta_-}g_i(\theta)d\theta,\quad
\hat{G}_i=M\int^{\pi/2}_{\theta_o}g_i(\theta)d\theta,
\ee
and
\be
g_{\theta}=\frac{1}{\sqrt{\Theta(\theta)}},\quad
g_{\phi}=\frac{\csc^2\theta}{\sqrt{\Theta(\theta)}},\quad
g_{t}=\frac{\cos^2\theta}{\sqrt{\Theta(\theta)}},
\ee
where $\Theta(\theta)$ is the angular potential defined in \eqref{eq:NullAngularPoten}
and $\theta_{\pm}$ are roots of it.

Following Refs~\cite{Gralla:2017ufe,Kapec:2019hro}, we obtain the results for these integrals in the near-extremal limit, as
\begin{subequations}
\bea
G_{\theta}&=&\frac{2}{(1-\tilde{b}^2)\sqrt{-\mathcal{I}_-}}K\Bigg(
\frac{\mathcal{I}_+}{\mathcal{I}_-}\Bigg)+\mathcal{O}(\epsilon),\\
\hat{G}_{\theta}&=&\frac{1}{(1-\tilde{b}^2)\sqrt{-\mathcal{I}_-}}F\Bigg(\Psi_o\Big|
\frac{\mathcal{I}_+}{\mathcal{I}_-}\Bigg)+\mathcal{O}(\epsilon),\\
G_{\phi}&=&\frac{2}{(1-\tilde{b}^2)\sqrt{-\mathcal{I}_-}}\Pi\Bigg(\mathcal{I}_+\Big|
\frac{\mathcal{I}_+}{\mathcal{I}_-}\Bigg)+\mathcal{O}(\epsilon),\\
\hat{G}_{\phi}&=&\frac{1}{(1-\tilde{b}^2)\sqrt{-\mathcal{I}_-}}\Pi\Bigg(\mathcal{I}_+;\Psi_o\Big|
\frac{\mathcal{I}_+}{\mathcal{I}_-}\Bigg)+\mathcal{O}(\epsilon),\\
G_{t}&=&-\frac{4\mathcal{I}_+}{(1-\tilde{b}^2)\sqrt{-\mathcal{I}_-}}E^{\prime}\Bigg(
\frac{\mathcal{I}_+}{\mathcal{I}_-}\Bigg)+\mathcal{O}(\epsilon),\\
\hat{G}_{t}&=&-\frac{2\mathcal{I}_+}{(1-\tilde{b}^2)\sqrt{-\mathcal{I}_-}}
E^{\prime}\Bigg(\Psi_o\Big|
\frac{\mathcal{I}_+}{\mathcal{I}_-}\Bigg)+\mathcal{O}(\epsilon),
\eea
\end{subequations}
where
\be
\mathcal{I}_{\pm}=\frac{q^2-6+2\tilde{b}^2\pm\sqrt{(4-q^2)(12-q^2-8\tilde{b})}}
{2(1-\tilde{b}^2)},
\ee
and
\be
\Psi_o=\arcsin\sqrt{\frac{\cos^2\theta_o}{\mathcal{I}_+}} \ . 
\ee
In addition, $E^{\prime}(\phi|m)=\partial_m E(\phi|m)$, and $F(\phi|m)$, $E(\phi|m)$, $\Pi(n;\phi|m)$ are the incomplete elliptic integrals of the first, second and third kind, respectively, and $K(m)$, $E(m)$, $\Pi(n|m)$ are the complete elliptic integrals of the first, second and third kind, respectively.


\begin{thebibliography}{10}

\bibitem{Abbott:2016blz}
{\bfseries LIGO Scientific, Virgo} Collaboration, B.~P. Abbott {\em et~al.},
  ``{Observation of Gravitational Waves from a Binary Black Hole Merger},''
  \href{http://dx.doi.org/10.1103/PhysRevLett.116.061102}{{\em Phys. Rev.
  Lett.} {\bfseries 116} no.~6, (2016) 061102},
\href{http://arxiv.org/abs/1602.03837}{{\ttfamily arXiv:1602.03837 [gr-qc]}}.

\bibitem{Abbott:2016nmj}
{\bfseries LIGO Scientific, Virgo} Collaboration, B.~P. Abbott {\em et~al.},
  ``{GW151226: Observation of Gravitational Waves from a 22-Solar-Mass Binary
  Black Hole Coalescence},''
  \href{http://dx.doi.org/10.1103/PhysRevLett.116.241103}{{\em Phys. Rev.
  Lett.} {\bfseries 116} no.~24, (2016) 241103},
\href{http://arxiv.org/abs/1606.04855}{{\ttfamily arXiv:1606.04855 [gr-qc]}}.

\bibitem{Akiyama:2019cqa}
{\bfseries Event Horizon Telescope} Collaboration, K.~Akiyama {\em et~al.},
  ``{First M87 Event Horizon Telescope Results. I. The Shadow of the
  Supermassive Black Hole},''
  \href{http://dx.doi.org/10.3847/2041-8213/ab0ec7}{{\em Astrophys. J.}
  {\bfseries 875} no.~1, (2019) L1},
\href{http://arxiv.org/abs/1906.11238}{{\ttfamily arXiv:1906.11238
  [astro-ph.GA]}}.

\bibitem{Akiyama:2019brx}
{\bfseries Event Horizon Telescope} Collaboration, K.~Akiyama {\em et~al.},
  ``{First M87 Event Horizon Telescope Results. II. Array and
  Instrumentation},'' \href{http://dx.doi.org/10.3847/2041-8213/ab0c96}{{\em
  Astrophys. J.} {\bfseries 875} no.~1, (2019) L2},
\href{http://arxiv.org/abs/1906.11239}{{\ttfamily arXiv:1906.11239
  [astro-ph.IM]}}.

\bibitem{Akiyama:2019sww}
{\bfseries Event Horizon Telescope} Collaboration, K.~Akiyama {\em et~al.},
  ``{First M87 Event Horizon Telescope Results. III. Data Processing and
  Calibration},'' \href{http://dx.doi.org/10.3847/2041-8213/ab0c57}{{\em
  Astrophys. J.} {\bfseries 875} no.~1, (2019) L3},
\href{http://arxiv.org/abs/1906.11240}{{\ttfamily arXiv:1906.11240
  [astro-ph.GA]}}.

\bibitem{Akiyama:2019bqs}
{\bfseries Event Horizon Telescope} Collaboration, K.~Akiyama {\em et~al.},
  ``{First M87 Event Horizon Telescope Results. IV. Imaging the Central
  Supermassive Black Hole},''
  \href{http://dx.doi.org/10.3847/2041-8213/ab0e85}{{\em Astrophys. J.}
  {\bfseries 875} no.~1, (2019) L4},
\href{http://arxiv.org/abs/1906.11241}{{\ttfamily arXiv:1906.11241
  [astro-ph.GA]}}.

\bibitem{Akiyama:2019fyp}
{\bfseries Event Horizon Telescope} Collaboration, K.~Akiyama {\em et~al.},
  ``{First M87 Event Horizon Telescope Results. V. Physical Origin of the
  Asymmetric Ring},'' \href{http://dx.doi.org/10.3847/2041-8213/ab0f43}{{\em
  Astrophys. J.} {\bfseries 875} no.~1, (2019) L5},
\href{http://arxiv.org/abs/1906.11242}{{\ttfamily arXiv:1906.11242
  [astro-ph.GA]}}.

\bibitem{Akiyama:2019eap}
{\bfseries Event Horizon Telescope} Collaboration, K.~Akiyama {\em et~al.},
  ``{First M87 Event Horizon Telescope Results. VI. The Shadow and Mass of the
  Central Black Hole},'' \href{http://dx.doi.org/10.3847/2041-8213/ab1141}{{\em
  Astrophys. J.} {\bfseries 875} no.~1, (2019) L6},
\href{http://arxiv.org/abs/1906.11243}{{\ttfamily arXiv:1906.11243
  [astro-ph.GA]}}.

\bibitem{Barack:2018yly}
L.~Barack {\em et~al.}, ``{Black holes, gravitational waves and fundamental
  physics: a roadmap},'' \href{http://dx.doi.org/10.1088/1361-6382/ab0587}{{\em
  Class. Quant. Grav.} {\bfseries 36} no.~14, (2019) 143001},
\href{http://arxiv.org/abs/1806.05195}{{\ttfamily arXiv:1806.05195 [gr-qc]}}.

\bibitem{Cunha:2018acu}
P.~V.~P. Cunha and C.~A.~R. Herdeiro, ``{Shadows and strong gravitational
  lensing: a brief review},''
  \href{http://dx.doi.org/10.1007/s10714-018-2361-9}{{\em Gen. Rel. Grav.}
  {\bfseries 50} no.~4, (2018) 42},
\href{http://arxiv.org/abs/1801.00860}{{\ttfamily arXiv:1801.00860 [gr-qc]}}.

\bibitem{Gralla:2019xty}
S.~E. Gralla, D.~E. Holz, and R.~M. Wald, ``{Black Hole Shadows, Photon Rings,
  and Lensing Rings},''
  \href{http://dx.doi.org/10.1103/PhysRevD.100.024018}{{\em Phys. Rev.}
  {\bfseries D100} no.~2, (2019) 024018},
\href{http://arxiv.org/abs/1906.00873}{{\ttfamily arXiv:1906.00873
  [astro-ph.HE]}}.

\bibitem{Johnson:2019ljv}
M.~D. Johnson {\em et~al.}, ``{Universal Interferometric Signatures of a Black
  Hole's Photon Ring},''
\href{http://arxiv.org/abs/1907.04329}{{\ttfamily arXiv:1907.04329
  [astro-ph.IM]}}.

\bibitem{Bambi:2019testing}
C.~Bambi, K.~Freese, S.~Vagnozzi, and L.~Visinelli, ``Testing the rotational nature of the supermassive object M87* from the circularity and size of its first image"
\href{https://journals.aps.org/prd/abstract/10.1103/PhysRevD.100.044057}{{\em Phys. Rev.}
  {\bfseries D100} (2019) 044057}.

\bibitem{cunningham1972optical}
C.~Cunningham and J.~Bardeen, ``The optical appearance of a star orbiting an
  extreme kerr black hole,''
\href{http://dx.doi.org/10.1086/180933}{{\em The Astrophysical Journal}
  {\bfseries 173} (1972) L137}.

\bibitem{cunningham1973optical}
C.~Cunningham and J.~M. Bardeen, ``The optical appearance of a star orbiting an
  extreme kerr black hole,'' \href{http://dx.doi.org/10.1086/152223}{{\em The
  Astrophysical Journal} {\bfseries 183} (1973) 237--264}.

\bibitem{Gralla:2017ufe}
S.~E. Gralla, A.~Lupsasca, and A.~Strominger, ``{Observational Signature of
  High Spin at the Event Horizon Telescope},''
  \href{http://dx.doi.org/10.1093/mnras/sty039}{{\em Mon. Not. Roy. Astron.
  Soc.} {\bfseries 475} no.~3, (2018) 3829--3853},
\href{http://arxiv.org/abs/1710.11112}{{\ttfamily arXiv:1710.11112
  [astro-ph.HE]}}.

\bibitem{Yan:2019etp}
H.~Yan, ``{Influence of a plasma on the observational signature of a high-spin
  Kerr black hole},'' \href{http://dx.doi.org/10.1103/PhysRevD.99.084050}{{\em
  Phys. Rev.} {\bfseries D99} no.~8, (2019) 084050},
\href{http://arxiv.org/abs/1903.04382}{{\ttfamily arXiv:1903.04382 [gr-qc]}}.

\bibitem{Dokuchaev:2018kzk}
V.~I. Dokuchaev and N.~O. Nazarova, ``{Event horizon image within black hole
  shadow},'' \href{http://dx.doi.org/10.1134/S1063776119030026}{{\em J. Exp.
  Theor. Phys.} {\bfseries 128} no.~4, (2019) 578--585},
\href{http://arxiv.org/abs/1804.08030}{{\ttfamily arXiv:1804.08030
  [astro-ph.HE]}}.

\bibitem{Dokuchaev:2019bbf}
V.~I. Dokuchaev, N.~O. Nazarova, and V.~P. Smirnov, ``{Event horizon
  silhouette: implications to supermassive black holes in the galaxies M87 and
  Milky Way},'' \href{http://dx.doi.org/10.1007/s10714-019-2564-8}{{\em Gen.
  Rel. Grav.} {\bfseries 51} no.~6, (2019) 81},
\href{http://arxiv.org/abs/1903.09594}{{\ttfamily arXiv:1903.09594
  [astro-ph.HE]}}.

\bibitem{Dokuchaev:2019pcx}
V.~I. Dokuchaev and N.~O. Nazarova, ``{The brightest point in accretion disk
  and black hole spin: implication to the image of black hole M87*},''
  \href{http://dx.doi.org/10.3390/universe5080183}{{\em Universe} {\bfseries 5}
  (2019) 183},
\href{http://arxiv.org/abs/1906.07171}{{\ttfamily arXiv:1906.07171
  [astro-ph.HE]}}.

\bibitem{sen1992rotating}
A.~Sen, ``{Rotating charged black hole solution in heterotic string theory},''
  \href{http://dx.doi.org/10.1103/PhysRevLett.69.1006}{{\em Phys. Rev. Lett.}
  {\bfseries 69} (1992) 1006--1009},
\href{http://arxiv.org/abs/hep-th/9204046}{{\ttfamily arXiv:hep-th/9204046
  [hep-th]}}.

\bibitem{ayzenberg2014slowly}
D.~Ayzenberg and N.~Yunes, ``{Slowly-Rotating Black Holes in
  Einstein-Dilaton-Gauss-Bonnet Gravity: Quadratic Order in Spin Solutions},''
  \href{http://dx.doi.org/10.1103/PhysRevD.91.069905,
  10.1103/PhysRevD.90.044066}{{\em Phys. Rev.} {\bfseries D90} (2014) 044066},
  \href{http://arxiv.org/abs/1405.2133}{{\ttfamily arXiv:1405.2133 [gr-qc]}}.
[Erratum: Phys. Rev.D91,no.6,069905(2015)].

\bibitem{myers1986black}
R.~C. Myers and M.~J. Perry, ``{Black Holes in Higher Dimensional
  Space-Times},''
\href{http://dx.doi.org/10.1016/0003-4916(86)90186-7}{{\em Annals Phys.}
  {\bfseries 172} (1986) 304}.

\bibitem{amarilla2010null}
L.~Amarilla, E.~F. Eiroa, and G.~Giribet, ``{Null geodesics and shadow of a
  rotating black hole in extended Chern-Simons modified gravity},''
  \href{http://dx.doi.org/10.1103/PhysRevD.81.124045}{{\em Phys. Rev.}
  {\bfseries D81} (2010) 124045},
\href{http://arxiv.org/abs/1005.0607}{{\ttfamily arXiv:1005.0607 [gr-qc]}}.

\bibitem{abdujabbarov2013shadow}
A.~Abdujabbarov, F.~Atamurotov, Y.~Kucukakca, B.~Ahmedov, and U.~Camci,
  ``{Shadow of Kerr-Taub-NUT black hole},''
  \href{http://dx.doi.org/10.1007/s10509-012-1337-6}{{\em Astrophys. Space
  Sci.} {\bfseries 344} (2013) 429--435},
\href{http://arxiv.org/abs/1212.4949}{{\ttfamily arXiv:1212.4949
  [physics.gen-ph]}}.

\bibitem{moffat2006scalar}
J.~W. Moffat, ``{Scalar-tensor-vector gravity theory},''
  \href{http://dx.doi.org/10.1088/1475-7516/2006/03/004}{{\em JCAP} {\bfseries
  0603} (2006) 004},
\href{http://arxiv.org/abs/gr-qc/0506021}{{\ttfamily arXiv:gr-qc/0506021
  [gr-qc]}}.

\bibitem{Guo:2018kis}
M.~Guo, N.~A. Obers, and H.~Yan, ``{Observational signatures of near-extremal
  Kerr-like black holes in a modified gravity theory at the Event Horizon
  Telescope},'' \href{http://dx.doi.org/10.1103/PhysRevD.98.084063}{{\em Phys.
  Rev.} {\bfseries D98} no.~8, (2018) 084063},
\href{http://arxiv.org/abs/1806.05249}{{\ttfamily arXiv:1806.05249 [gr-qc]}}.

\bibitem{An:2017hby}
J.~An, J.~Peng, Y.~Liu, and X.-H. Feng, ``{Kerr-Sen Black Hole as Accelerator
  for Spinning Particles},''
  \href{http://dx.doi.org/10.1103/PhysRevD.97.024003}{{\em Phys. Rev.}
  {\bfseries D97} no.~2, (2018) 024003},
\href{http://arxiv.org/abs/1710.08630}{{\ttfamily arXiv:1710.08630 [gr-qc]}}.

\bibitem{Gyulchev:2006zg}
G.~N. Gyulchev and S.~S. Yazadjiev, ``{Kerr-Sen dilaton-axion black hole
  lensing in the strong deflection limit},''
  \href{http://dx.doi.org/10.1103/PhysRevD.75.023006}{{\em Phys. Rev.}
  {\bfseries D75} (2007) 023006},
\href{http://arxiv.org/abs/gr-qc/0611110}{{\ttfamily arXiv:gr-qc/0611110
  [gr-qc]}}.

\bibitem{Younsi:2016azx}
Z.~Younsi, A.~Zhidenko, L.~Rezzolla, R.~Konoplya, and Y.~Mizuno, ``{New method
  for shadow calculations: Application to parametrized axisymmetric black
  holes},'' \href{http://dx.doi.org/10.1103/PhysRevD.94.084025}{{\em Phys.
  Rev.} {\bfseries D94} no.~8, (2016) 084025},
\href{http://arxiv.org/abs/1607.05767}{{\ttfamily arXiv:1607.05767 [gr-qc]}}.

\bibitem{PhysRevD.78.044007}
K.~Hioki and U.~Miyamoto, ``Hidden symmetries, null geodesics, and photon
  capture in the sen black hole,''
  \href{http://dx.doi.org/10.1103/PhysRevD.78.044007}{{\em Phys. Rev. D}
  {\bfseries 78} (Aug, 2008) 044007}.
  \url{https://link.aps.org/doi/10.1103/PhysRevD.78.044007}.

\bibitem{Dastan:2016bfy}
S.~Dastan, R.~Saffari, and S.~Soroushfar, ``{Shadow of a Kerr-Sen dilaton-axion
  Black Hole},''
\href{http://arxiv.org/abs/1610.09477}{{\ttfamily arXiv:1610.09477 [gr-qc]}}.

\bibitem{Uniyal:2017yll}
R.~Uniyal, H.~Nandan, and K.~D. Purohit, ``{Null geodesics and observables
  around the Kerr-Sen black hole},''
  \href{http://dx.doi.org/10.1088/1361-6382/aa9ad9}{{\em Class. Quant. Grav.}
  {\bfseries 35} no.~2, (2018) 025003},
\href{http://arxiv.org/abs/1703.07510}{{\ttfamily arXiv:1703.07510 [gr-qc]}}.

\bibitem{carter1968global}
B.~Carter, ``{Global structure of the Kerr family of gravitational fields},''
\href{http://dx.doi.org/10.1103/PhysRev.174.1559}{{\em Phys. Rev.} {\bfseries
  174} (1968) 1559--1571}.

\bibitem{bardeen1972rotating}
J.~M. Bardeen, W.~H. Press, and S.~A. Teukolsky, ``{Rotating black holes:
  Locally nonrotating frames, energy extraction, and scalar synchrotron
  radiation},''
\href{http://dx.doi.org/10.1086/151796}{{\em Astrophys. J.} {\bfseries 178}
  (1972) 347}.

\bibitem{porfyriadis2017photon}
A.~P. Porfyriadis, Y.~Shi, and A.~Strominger, ``{Photon Emission Near Extreme
  Kerr Black Holes},'' \href{http://dx.doi.org/10.1103/PhysRevD.95.064009}{{\em
  Phys. Rev.} {\bfseries D95} no.~6, (2017) 064009},
\href{http://arxiv.org/abs/1607.06028}{{\ttfamily arXiv:1607.06028 [gr-qc]}}.

\bibitem{carlson20153}
E.~Carlson, T.~Jeltema, and S.~Profumo, ``{Where do the 3.5 keV photons come
  from? A morphological study of the Galactic Center and of Perseus},''
  \href{http://dx.doi.org/10.1088/1475-7516/2015/02/009}{{\em JCAP} {\bfseries
  1502} no.~02, (2015) 009},
\href{http://arxiv.org/abs/1411.1758}{{\ttfamily arXiv:1411.1758
  [astro-ph.HE]}}.

\bibitem{sheoran2017mass}
P.~Sheoran, A.~Herrera-Aguilar, and U.~Nucamendi, ``{Mass and spin of a
  Kerr-MOG black hole and a test for the Kerr black hole hypothesis},''
\href{http://arxiv.org/abs/1712.03344}{{\ttfamily arXiv:1712.03344 [gr-qc]}}.

\bibitem{Guo:2019photon}
M.~Guo, P.-C. Li, and B.~Chen, ``{Photon emission near Myers-Perry black holes
  in the large dimension limit},''
  \href{https://arxiv.org/abs/1911.08814}{{\ttfamily arXiv:1911.08814 [gr-qc]}}.

\bibitem{Kapec:2019hro}
D.~Kapec and A.~Lupsasca, ``{Particle motion near high-spin black holes},''
\href{http://arxiv.org/abs/1905.11406}{{\ttfamily arXiv:1905.11406 [hep-th]}}.

\end{thebibliography}
\providecommand{\href}[2]{#2}\begingroup\raggedright\endgroup

\end{document}